\pdfoutput=1 % only if pdf/png/jpg images are used
\documentclass{JINST}

\title{Simulation of the Dynamic Inefficiency of the CMS Pixel Detector}

\author{M\'arton Bart\'ok$^{a,b}$ for the CMS Collaboration\\
\llap{$^a$}University of Debrecen,\\
  H-4032, 1. Egyetem t\'er, HUNGARY\\
\llap{$^b$}Wigner Research Centre for Physics,\\
  H-1121, 29-33. Konkoly Thege Mikl\'os \'ut, Budapest, HUNGARY\\
E-mail: \email{marton.bartok@cern.ch}}

\abstract{The Pixel Detector is the innermost part of the CMS Tracker. It therefore has to prevail in the harshest environment in terms of particle fluence and radiation. There are several mechanisms that may decrease the efficiency of the detector. These are mainly caused by data acquisition (DAQ) problems and/or Single Event Upsets (SEU). Any remaining efficiency loss is referred to as the dynamic inefficiency. It is caused by various mechanisms inside the Readout Chip (ROC) and depends strongly on the data occupancy. In the 2012 data, at high values of instantaneous luminosity the inefficiency reached 2\% (in the region closest to the interaction point) which is not negligible. In the 2015 run higher instantaneous luminosity is expected, which will result in lower efficiencies; therefore this effect needs to be understood and simulated. A data-driven method has been developed to simulate dynamic inefficiency, which has been shown to successfully simulate the effects.}

\keywords{CMS collaboration; LHC; Pixel Detector; Dynamic Inefficiency}

\begin{document}

\section{Introduction}\label{sec:intro}

\begin{figure}[tbp]
\centering
\includegraphics[width=1\textwidth]{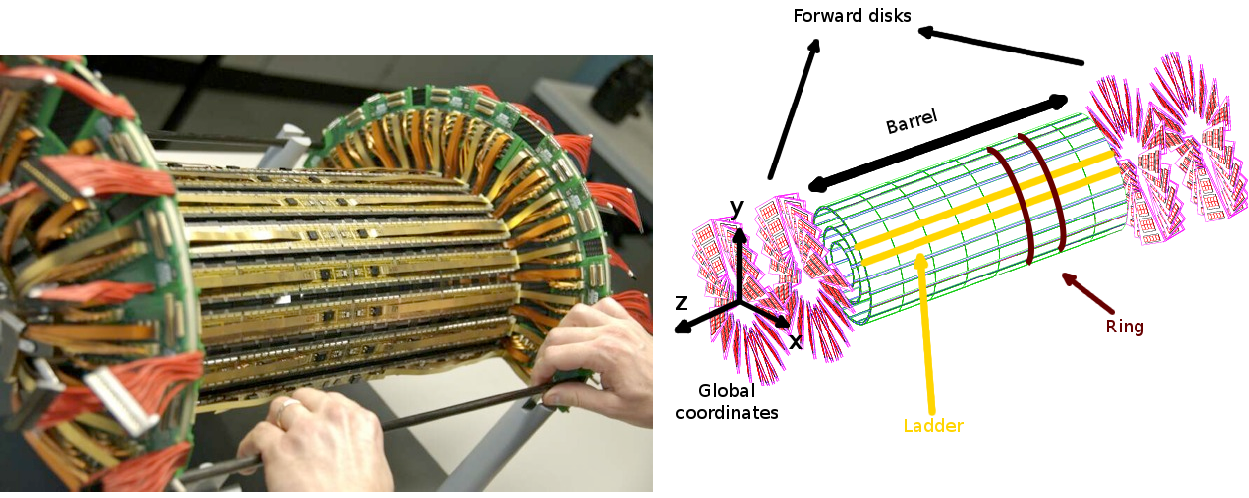}
\caption{The structure of the CMS Pixel Detector.}
\label{fig:pix_det}
\end{figure}

\begin{figure}[tbp]
\centering
\includegraphics[width=.4\textwidth]{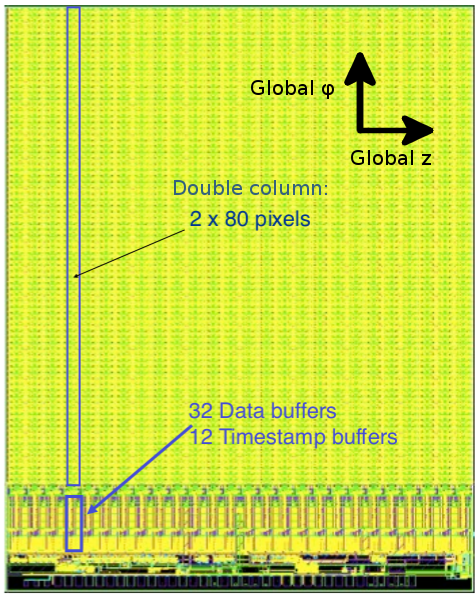}
\caption{Schematics of a Pixel Readout Chip (ROC) with the global coordinates relevant for the barrel part of the detector indicated.}
\label{fig:ROC}
\end{figure}

The Compact Muon Solenoid (CMS) is one of the two general-purpose detectors that measure the products of high energy particle interactions at the Large Hadron Collider (LHC). The CMS Pixel Detector is a silicon semiconductor detector at the centre of the CMS tracking system. Along with the surrounding Silicon Strip Tracker, it provides precision measurements of the trajectories of charged particles. The Pixel Detector consist of three cylindrical layers called the barrel and two endcap disks at each end, called the forward part of the detector. It lies very close to the interaction point; the mean radius of layers 1, 2 and 3 are 4.4 cm, 7.3 cm and 10.2 cm, respectively. The barrel is divided into ladders (along the $r-\phi$ plane) and rings (along the $z$ axis) as can be seen in figure~\ref{fig:pix_det}. The intersection of a ladder and a ring is called a module, the basic building block of the detector. A module is made up of 8 or 16 Readout Chips (ROC), each with 52$\times$80 pixels of size 100$\times$150 $\mu m^2$. The ROC reads out the pixel data in double columns, with each double column having its own data and time-stamp buffer~\cite{double_column}. The structure of a ROC can be seen in figure~\ref{fig:ROC}. A more detailed description of the detector can be found in~\cite{CMS}.

\section{Dynamic Inefficiency}

We measure the performance of the detector in terms of hit efficiency~\cite{hit_eff}. The hit efficiency is the probability to find a pixel cluster in any given sensor within a $500\mu m$ radius of a charged particle trajectory. Only well reconstructed tracks are used in the measurement. Tracks are required to be isolated from other tracks, to originate from the primary vertex, and to pass through the active regions of the sensors. We estimate the systematic uncertainty for the measurement by comparing it to measurements with ``ideal tracks'' (high transverse momentum, zero impact parameter, etc.). This way the systematic uncertainty is approximated to be $0.3\%$. This method is used to measure hit efficiency both in the data and simulations. The sensor efficiency is defined by excluding the following sources of inefficiency: permanently damaged detector parts, modules with readout errors and Single Event Upsets (SEU). The last one is caused by ionising radiation, which can cause the memory state of a logical element of the detector to flip. This may affect individual pixels, ROCs, or the readout electronics for entire modules. The impact of SEUs is minimized by reprogramming the detector during data taking~\cite{SEU}. ROCs undergoing SEU in recorded data (taken before the reprogramming happens) are identified and excluded from the hit efficiency measurement. The SEU is not yet included in the CMS simulation software since it has a minor effect on recorded data. Excluding all the above efficiency decreasing effects from the measurement, there remains a significant efficiency loss that we call the dynamic inefficiency. In a series of high multiplicity events, the buffers (mainly data and time-stamp buffers) of the ROC may overflow, resulting in data losses. The largest part of the inefficiency comes from the overflow of the time-stamp buffers in the double columns. After a triggered readout, the buffers get cleared and the overflowed buffers are able to store data again. Therefore we call this inefficiency dynamic in the sense that it is not permanent, instead the size of the effect depends on trigger rate, previous and current events. The detector has a maximum efficiency just after the abort gap of the LHC. Individual pixels and entire ROCs can be inefficient, but data suggests that double column loss is the dominant effect.

\section{Simulation}

The dynamic inefficiency in a certain event depends on pixel occupancy in the double columns in previous events. In order to properly simulate this effect, one would need to use the full simulation of the ROCs, in addition to storing the history of many events. Neither of this is possible in the current CMS simulation software. The average occupancy is determined by the pileup in a series of events. The simulation is generated with a flat pileup distribution in a range that covers the entire domain of pileup expected in the data. After measuring the real pileup in the data, the pileup distribution of the simulation is re-weighted accordingly. When the simulation is created, the final pileup distribution is unknown, thus the occupancy could not be calculated but for one event, which is inadequate. Therefore, a data-driven method has been developed, in which the hit efficiency is parametrised for each module as a function of variables determining occupancy: pileup and module position (layer, ladder and ring coordinates). This way, the simulated efficiency is also independent of the method of the pileup simulation, but it has to be calibrated for different running conditions, such as the bunchspacing and the energy of the collision.

The dynamic inefficiency is observed in the hit efficiency measurement of the detector, and it is expected to be caused by double column loss. Therefore in the simulation we set the double column efficiency in such a way that the measured hit efficiency would be the same in the data as in the simulation. In order to do that, first a double column efficiency scan is made: a series of simulated datasets are created with different double column efficiency settings. By measuring the hit efficiency in each simulation, one can derive a hit efficiency to double column efficiency conversion function. With the help of this function and the hit efficiency measured in the data, scaling factors can be derived for each module as an input for the simulation. The dynamic inefficiency is simulated in this way, in all three layers of the barrel part of the detector.

Module position is determined by ladder coordinates, ring coordinates and layer number. The ladder and ring coordinates are defined in the CMS global coordinate system: $x$ points towards the centre of the LHC ring, $y$ points upwards to the surface, and $z$ points along the beam line. This can be seen in figure~\ref{fig:pix_det}. The $x=y=z=0$ is the interaction point in the centre of the CMS detector. The ladders are numbered along the $r-\phi$ plane. The coordinates' sign corresponds to the $x$ axis sign; the numbers start from $x=0$ and run from the $+y$ side to the $-y$ side. The rings are numbered along the beam axis with the interaction point in the middle. The coordinates' sign corresponds to the $z$ axis sign; the numbers increase in magnitude moving outwards in $\pm z$.

\begin{figure}[tbp]
\centering
\includegraphics[width=1\textwidth]{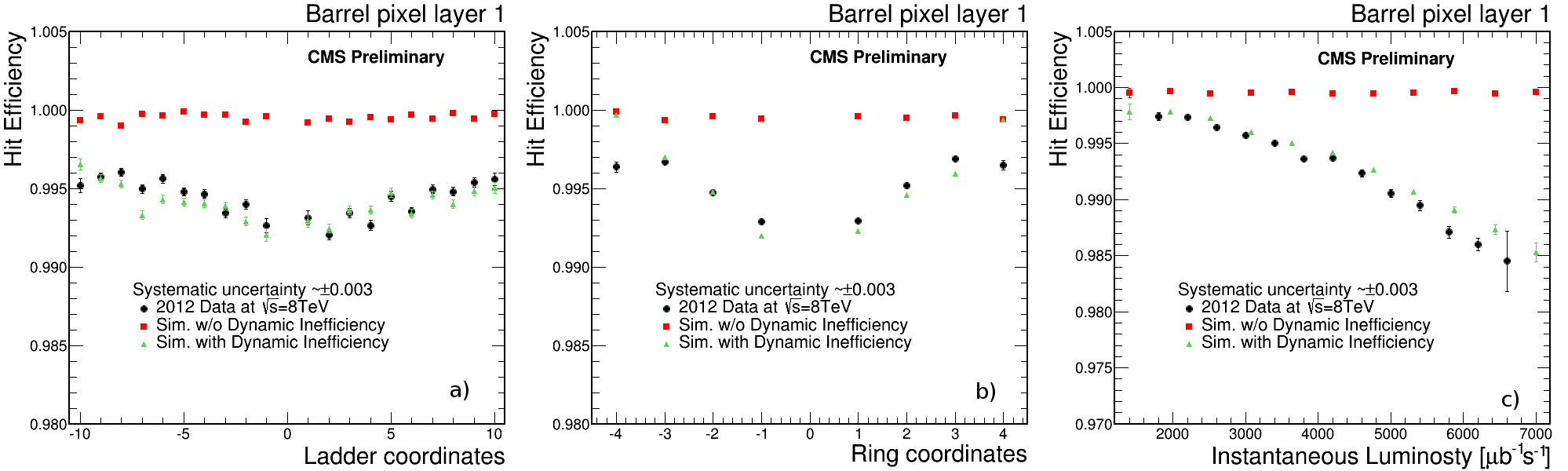}
\caption{Hit efficiency as a function of the three variables that the dynamic inefficiency simulation is parameterised in terms of: \textbf{a)} ladder coordinates, \textbf{b)} ring coordinates and \textbf{c)} instantaneous luminosity.}
\label{fig:dyn_ineff}
\end{figure}

Results of the simulation are shown for layer 1 where the effect is most visible. The method has been validated by comparing simulations with, and without, dynamic inefficiencies to data. The data showed in the following plots are from the 2012 $8TeV$ data taking period used with ZeroBias trigger and with an average pileup of $17.72$. Pileup is not directly observed in the data, it is converted from the instantaneous luminosity measurement, this conversion can also be done for the simulation in the other way around. In figure~\ref{fig:dyn_ineff}, the dynamic inefficiency simulation agrees with the data by construction, as the double column efficiency in the simulation was set to reproduce hit efficiency in data. Ladder coordinates correspond to azimuthal angle $\phi$, in which the detector is symmetric. However because the beam was not perfectly centered (beam offset) there is an azimuthal asymmetry in figure~\ref{fig:dyn_ineff}(a). In figure~\ref{fig:dyn_ineff}(b), the simulated hit efficiency measured in the rings numbered $\pm4$ (corresponding to the rings at each end of the barrel region) differ from those observed in the data, but still in the range of the estimated systematic uncertainty. This subject is discussed in the next section. In figure~\ref{fig:dyn_ineff}(c) it can be seen that the dependence of the hit efficiency on the instantaneous luminosity is well reproduced in the simulation.

\section{Results}

\begin{figure}[tbp]
\centering
\includegraphics[width=1\textwidth]{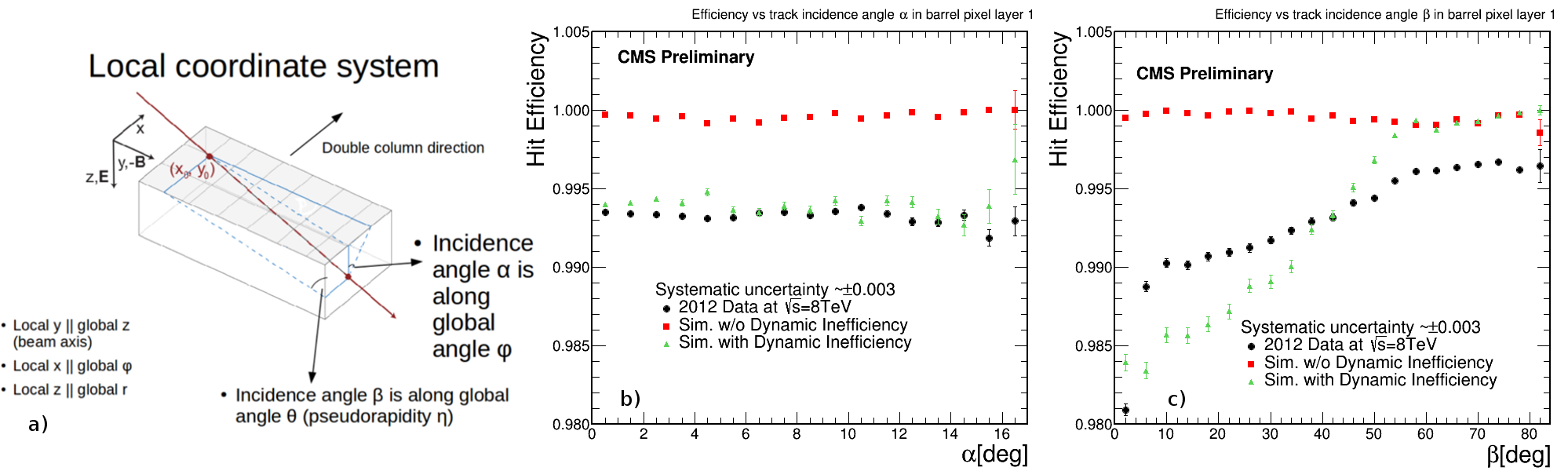}
\caption{\textbf{a)} Local coordinate system, \textbf{b)} Efficiency versus $\alpha$, \textbf{c)} Efficiency versus $\beta$.}
\label{fig:inc_angle}
\end{figure}

The assumption of double column loss causing the dynamic inefficiency can be verified by studying the track incidence angles. The incidence angles are defined in the local coordinate system, which can be seen in figure~\ref{fig:inc_angle}(a). The double column direction points along global azimuthal angle $\phi$, this can also be seen in figure~\ref{fig:ROC}. The incidence angle $\alpha$ points along the double column direction, whilst the incidence angle $\beta$ corresponds to the global polar angle $\theta$ (or equivalently, the pseudorapidity, $\eta$) which is perpendicular to the double columns.

Based on these definitions, we expect that if the dynamic inefficiency is caused by double column loss, the hit efficiency should be independent of $\alpha$, but dependent on $\beta$. This trend is indeed observed in figure~\ref{fig:inc_angle}. However, in figure~\ref{fig:inc_angle}(c) it can also be seen that the shape of the distribution in the simulation does not agree with that observed in the data. For values of $\beta$ close to $90^\circ$degrees, an incoming (grazing) track creates a long cluster and the loss of a double column will only have a small effect, because $\beta$ is perpendicular to the double column direction. Clusters cut in half by a double column loss will still most probably be matched to the track, therefore the hit efficiency is unchanged. The modules on ring numbers $\pm4$ (the rings at each end of the barrel region) are likely to have tracks with $\beta$ near $90^\circ$, therefore the hit efficiency measured there is mostly unaffected by double column loss. A track perpendicular to the module plane ($\beta \approx 0^\circ$) causes a small cluster which is more likely to lie within the boundary of a lost double column; as such the efficiency loss would be greater. While the difference between the simulation and the data is clear in figure~\ref{fig:inc_angle}(c), it is not significant if we take into account the systematic uncertainty which is estimated as $0.3\%$. The effect of the double column loss is possibly not well simulated, and could lead to this discrepancy, or the difference could suggest another efficiency decreasing mechanism. One should consider the fact that the maximum of the measured inefficiency is $\approx2\%$, the effect we try to measure is not that sizable enough to clearly see the underlying phenomenon.

\begin{figure}[tbp]
\centering
\includegraphics[width=1\textwidth]{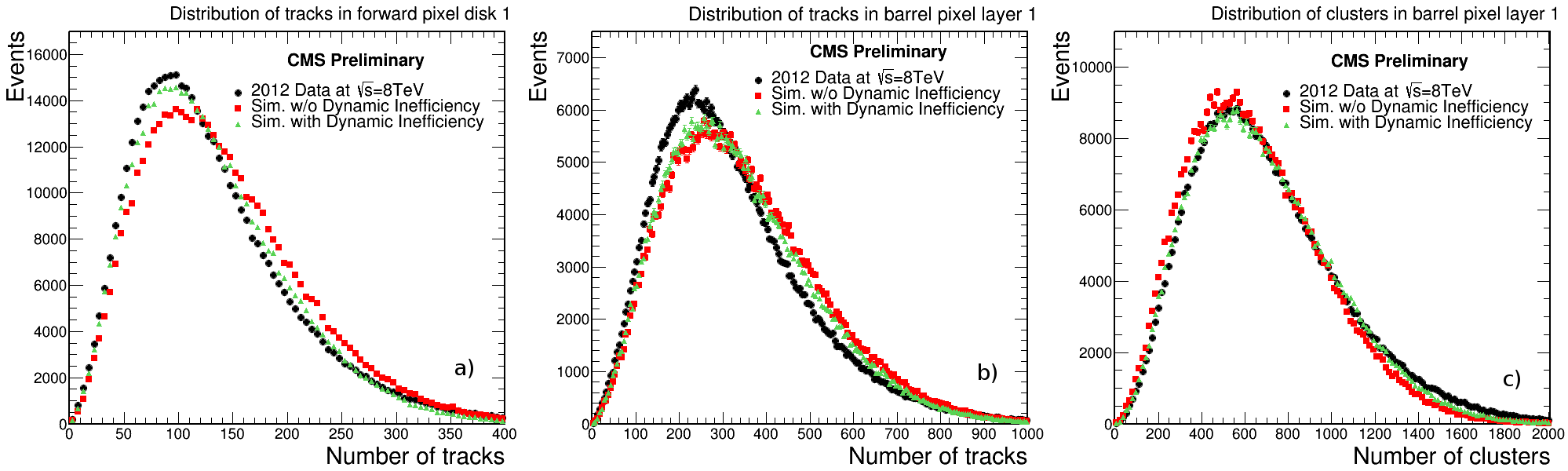}
\caption{The distribution of the number of tracks and clusters in the forward pixel detector and in layer 1 of the barrel pixel detector.}
\label{fig:ntrack_nclu}
\end{figure}

In order to cross check the validity of the method many parameters have been examined; the distribution of the number of tracks and clusters showed the most improvement. It is not expected that the distribution of these variables would be in perfect agreement with data solely by the inclusion of the dynamic inefficiency in the simulation. The signs of improvements implies the legitimacy of the way of the dynamic inefficiency is simulated. The results can be seen in figure~\ref{fig:ntrack_nclu}. The inefficiency of the forward disks is not simulated, however the effect of the barrel dynamic inefficiency simulation on the forward disk 1 can be seen in figure~\ref{fig:ntrack_nclu}(a). Tracks crossing the forward region of the detector are likely to have hits from the barrel region which serve as seeds for the tracking. Since these hits can be lost by dynamic inefficiency it affects the tracking in the forward region. The description of the distribution of the number of clusters has improved due to proper simulation of clusters splitted by double column loss.

\section{Conclusion}

The CMS simulation software has been improved by taking into account the dynamic inefficiency of the pixel detector barrel region. The advantage of using a data-driven method is that it is independent of the quality of the physics simulation. The disadvantage is that it needs to be calibrated using data from different running conditions. The technique has been validated by comparing several variables in data and simulation. The new simulation shows better agreement with data. Further improvement of the simulation is possible, for example, by including entire ROC or individual pixel loss and the extension of the efficiency loss to the forward disks.

\acknowledgments

The author wishes to thank to the Hungarian Scientific Research Fund (K 109703) for support.

\end{document}